\documentclass[10pt, conference, letterpaper]{IEEEtran}
\usepackage{cite}
\usepackage{amsmath,amssymb,amsfonts}
\usepackage{algorithmic}
\usepackage{graphicx}
\usepackage{textcomp}
\usepackage[usenames,dvipsnames,table]{xcolor} 
\usepackage{booktabs}
\usepackage{colortbl}
\usepackage{float}
\usepackage{tabularx}
\usepackage{subcaption}
\usepackage{url}
\PassOptionsToPackage{hyphens}{url}
\usepackage{xurl}
\usepackage[ruled,linesnumbered]{algorithm2e} 
\usepackage{pgfplots}
\usepackage{tikz}
\usetikzlibrary{shapes.geometric, arrows}
\pgfplotsset{compat=1.18} 
\usepackage{float}
\usepackage[square,numbers]{natbib}
\usepackage{hyperref}

\hypersetup{
    colorlinks=false,
    hidelinks
    }

\SetKwRepeat{Do}{do}{while}%

\def\BibTeX{{\rm B\kern-.05em{\sc i\kern-.025em b}\kern-.08em
    T\kern-.1667em\lower.7ex\hbox{E}\kern-.125emX}}

\begin{document}
\bstctlcite{IEEEexample:BSTcontrol}	
\sloppy
\newcommand{\al}[1]{{\color{green}{[al: #1]}}}
\newcommand{\tbw}[1]{{\color{blue}{tbw: #1}}}
\newcommand{\nc}[1]{{\color{orange}{nc: #1}}}

\title{RainCloud: Decentralized Coordination and Communication in Heterogeneous IoT Swarms\\
}

\makeatletter 
\newcommand{\linebreakand}{%
  \end{@IEEEauthorhalign}
  \hfill\mbox{}\par
  \mbox{}\hfill\begin{@IEEEauthorhalign}
}
\makeatother 

\author{
    \centering
    \IEEEauthorblockN{Filip Loisel} 
    \IEEEauthorblockA{
        DSG@TU Wien,
        AUXO GmbH\\
        Email: filip.loisel@auxo-software.com
    }
    \and
    \IEEEauthorblockN{Geri Zeqo} 
    \IEEEauthorblockA{
        DSG@TU Wien,
        AUXO GmbH\\
        Email: name.surname@auxo-software.com
    }
    \and
    \IEEEauthorblockN{Andrea Morichetta*}
    \IEEEauthorblockA{
        Distributed Systems Group, TU Wien\\        Email:a.morichetta@dsg.tuwien.ac.at
    }
    \linebreakand
    \IEEEauthorblockN{Anna Lackinger}
    \IEEEauthorblockA{
        Distributed Systems Group, TU Wien\\        Email:a.lackinger@dsg.tuwien.ac.at
    }
    \and
    \IEEEauthorblockN{Schahram Dustdar}
    \IEEEauthorblockA{
        Distributed Systems Group, TU Wien\\        Email:dustdar@dsg.tuwien.ac.at
    }
}


\IEEEoverridecommandlockouts
\IEEEpubid{\makebox[\columnwidth]{979-8-3503-4965-8/24/\$31.00~\copyright2024 IEEE \hfill}
\hspace{\columnsep}\makebox[\columnwidth]{ }}
\maketitle
\IEEEpubidadjcol
\begin{abstract}
The increasing volume and complexity of IoT systems demand a transition from the cloud-centric model to a decentralized IoT architecture in the so called Computing Continuum, with no or minimal reliance on central servers. This paradigm shift, however, raises novel research concerns for decentralized coordination, calling for accurate policies. However, building such strategies is not trivial.
Our work aims to relieve the DevOps engineers from this concern and propose a solution for autonomous, decentralized task allocation at runtime for IoT systems.
To this end, we present a semantic communication approach and an ad-hoc lightweight coordination strategy based on Ant Colony Optimization (ACO).
We compare the ACO strategy with Random Search and Gossip protocol-based algorithms. We conduct accurate experiments with up to a hundred nodes in both a static and a dynamic environment, i.e., with device outages. We show that ACO finds a matching node with the smallest hops and messages sent. While the Gossip strategy can allocate the most tasks successfully, ACO scales better, thus being a promising candidate for decentralized task coordination in IoT clusters.\end{abstract}


\section{Introduction}
\label{sec:intro}

Nowadays, IoT devices pervade several areas, from healthcare to urban scenarios, industrial settings. It is predicted that in 2025 41.6 billion devices will be in use.~\footnote{\url{https://blogs.idc.com/2021/01/06/future-of-industry-ecosystems-shared-data-and-insights/}} This sheer number of IoT devices and the massive amount of data generated poses significant challenges to the current cloud-centric~\cite{tsigkanos2019towards, firmani2024intend} IoT model: a centralized architecture where requests and data flow from IoT devices to the cloud is becoming a bottleneck, making it challenging to handle latency, increased required bandwidth, vulnerability to failures, privacy concerns, and inefficient use of resources.
Therefore, a paradigm shift must happen. As an alternative, Edge and Fog computing~\cite{villari2016osmotic} can provide relief to the central server, especially if collaborating in a unified computing continuum~\cite{morichetta2023intent, morichetta2024cohabitation}. Furthermore, IoT and Edge devices can be more self-reliant, i.e., self-managing entities that make autonomous decisions and using local resources. Still, failures can often occur; furthermore, in mobility, IoT nodes can arbitrarily leave and join the system. Therefore, IoT swarms~\cite{ferrer2021towards} are a viable alternative to the cloud-centric paradigm. Creating swarms for device collaboration means solving an optimization problem aiming at finding the best paths and edges across nodes. Typically, finding an optimal solution is complex; this task becomes even more challenging in real-world IoT environments with real-time constraints due to the heterogeneous and often battery-powered devices, the diverse requirements of the applications, which might not be supported by some devices, and the varying conditions of the network environment~\cite{xiao2022edge}.

\begin{figure}
    \centering
    \includegraphics[width=1.0\linewidth]{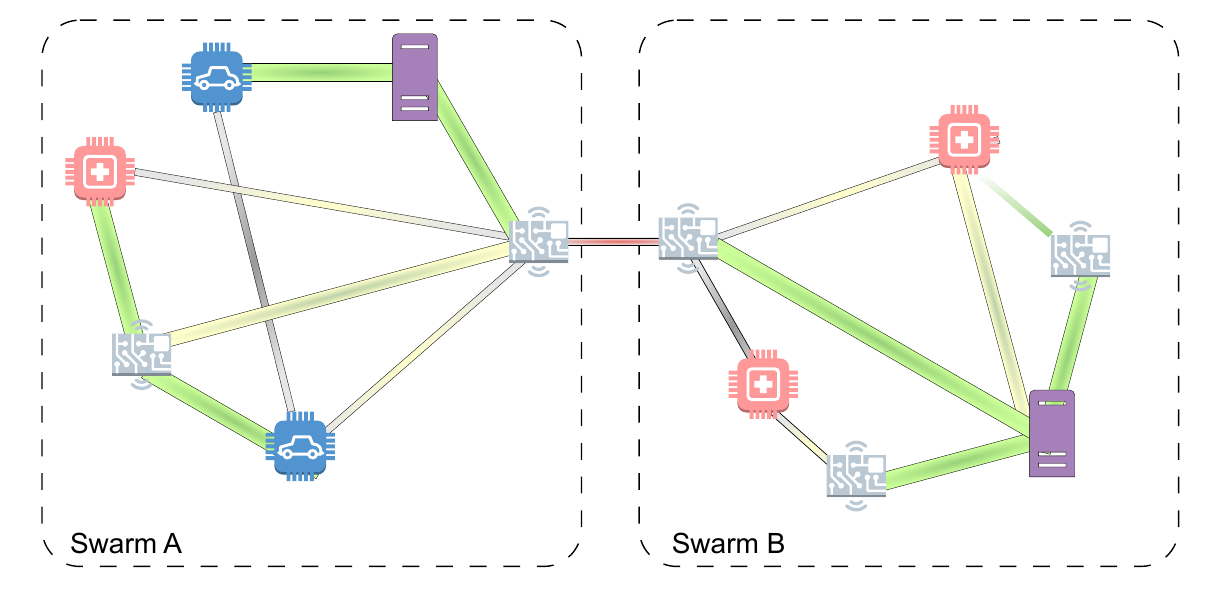}
    \caption{High-level representation of RainCloud's scenario.}
    \label{fig:highlevel-RainCloud}
\end{figure}

This paper addresses this challenge by examining the feasibility of a lightweight and real-time decentralized task coordination and communication mechanism for IoT devices. First, we design a node discovery and communication system that emphasizes the adaptation to the dynamic nature of a decentralized IoT environment. We achieve this by developing a framework, Rain Cloud System (RCS) that continuously adapts the network through constant checks, following the self-actualization paradigm~\cite{maslow1965self}. This framework leverages semantic metadata for message exchange and can adapt to different specific communication protocols. Second, we put forward a lightweight metaheuristic optimization algorithm based on Ant Colony Optimization (ACO) for task coordination, capable of finding effective paths and offering an efficient load distribution across the swarm by limiting the number of messages exchanged and the serving time. Furthermore, we focus the attention on providing a framework that can actually be used by infrastructure and system managers. Figure~\ref{fig:highlevel-RainCloud} illustrates these functionalities. The RainCloud System (RCS) facilitates the formation and updating of swarms in dynamic environments by continuously evaluating the quality of connections and optimal collaboration settings. Bolder and greener paths represent the most reliable edges, both in terms of connectivity and task offloading. Leveraging multiplatform technologies and languages, we provide an interoperable framework for decentralized coordination.
Therefore, through this paper, we wish to contribute meaningfully to decentralized task coordination in heterogeneous and dynamic IoT swarms, offering insights into efficient and reliable task coordination strategies and effective communication mechanisms. Furthermore, we are focusing on communication dynamics. Combined, we provide a holistic view and tools, including monitoring, for letting IoT devices function collaboratively in decentralized swarms.
Our contributions are as follows: 
\begin{itemize}
    \item We present a communication framework for decentralized coordination based on semantic information and self-actualization.
    \item We introduce a lightweight metaheuristic for swarm offloading based on Ant Colony concepts.
    \item The conduction and evaluation through a realistic simulation with up to 100 devices in a static environment and in a dynamic environment with device failures, offering a detailed comparison of our ACO-inspired coordination strategy with two other baselines: Random and Gossips.
    \item We offer a turnkey open-sourced Rain Cloud System (RCS)\footnote{\noindent
    \url{https://github.com/auxo-gmbh/rcs-flooding-scenario} \\
    \url{https://github.com/auxo-gmbh/rcs-documentation} \\
    \url{https://github.com/auxo-gmbh/rcs-monitoring} \\
    \url{⁠https://github.com/auxo-gmbh/rcs-rain-cloud-system}
    }
    framework for edge-based device connectivity.
\end{itemize}
\section{Related Work}
\label{sec:related}

In this Section, we explore the state of the art for decentralized task offloading in IoT scenarios with a focus on how other contributions handled communication in a decentralized system and we explore the use of variations of Ant Colony models for handling task offloading. We clarify how our approach goes beyond the state of the art by combining together these aspects in a unique, ready-to-deploy framework and draws away from previous research.


\subsection{Communication in decentralized IoT} \label{subsec-related:communication-IoT}
Communication plays an essential role in decentralized systems. Much effort has been put into exploring how to transmit information and what to consider as informative. The work of Bittman et al.~\cite{bittman2021don} focuses on the technology, offering a new Remote Procedure Call (RPC) approach in decentralized environments. The authors propose shifting from location-centric abstractions like RPC to data-centric abstractions akin to distributed shared memory (DSM) for better module composition. 
The work of Wang et al.~\cite{wang2021reference} also goes in the direction of RPC. It focuses on extending RPC with a shared address space and first-class references to promote interoperability and reduce task duplication. However, this approach shares the limit of the classic RPC model.
Kountouris et al.~\cite{kountouris2021semantics} propose a paradigm shift focused on the semantics of information, unifying information generation, transmission, and reconstruction. Semantic information spreading became the primary basis and paradigm for the RCS Framework Communication module with metadata information spreading. In particular, efforts toward semantic communication~\cite{seo2021semantics,uysal2022semantic,xie2020lite} seem to pave the path for richer communication patterns.

\subsection{Ant-Colony Optimization in decentralized coordination} \label{subsec-related:ACO_coordination}

In the context of large and dynamic environments, such as IoT, Kumar et al.~\cite{kumar2020comparative} propose a comparison between ACO and K-Means clustering algorithms for IoT job scheduling. These two approaches are studied to find the shortest route path, optimize QoS constraints, and reduce energy consumption. 
Wang et al.~\cite{wang2013ant} present the Collection Path Ant Colony Optimization (CPACO) method, which improves upon traditional parallel computing task allocation methods and ACO algorithms. They update the strategy in the Ant-Cycle Model, creating a three-dimensional path pheromone storage space. Similarly, Zannou et al.~\cite{zannou2019task} use ACO for task allocation, considering two parameters: the optimal path length and the nodes' capabilities. In distributed systems, ACO demonstrates the potential for improving QoS. Hussein and Mousa~\cite{hussein2020efficient} focus on reducing latency for delay-sensitive applications in IoT-Fog systems. 
In a similar fashion, Kishor and Chakarbarty~\cite{kishor2022task} propose a meta-heuristic scheduler called Smart Ant Colony Optimization (SACO) to offload tasks in a fog environment.
Latency is a key requirement for extremely dynamic applications, such as autonomous driving. Bui and Jung’s study~\cite{bui2019aco} explores the potential of ACO for dynamic decision-making in connected vehicles within an IoT environment through swarm intelligence. 
Tan et al.~\cite{tan2021energy} employ a bi-objective optimization approach with a fixed number of ants and a primary focus on energy efficiency and task completion time. Similarly, Wang et al.~\cite{wang2024bi} introduce a sophisticated bi-ACO framework, combining task offloading with UAV trajectory planning and using multiple heterogeneous colonies.
Xu et al.~\cite{xu2022research} present the GA-ACO Fusion Algorithm, proposing an algorithm that combines GA and ACO. This approach focuses on delay and energy efficiency.
Li et al.~\cite{li2023dqn} explore another combination, employing quantum computing principles to enhance the traditional ACO; through quantum bits, it expands the search space without increasing the number of ants. A quantum-boosted model is also presented by Dong et al.~\cite{dong2022quantum} even if without explicit ACO-based techniques.

\subsection{Summary}
Unlike the other approaches, our ACO implementation offers a fast and lightweight solution for real-time optimization of multiple variables, as well as efficient queue size management. This approach is suitable for real-world and dynamic IoT and edge computing scenarios, thanks to its fast optimization through reducing the number of ants and carrying the workload during the exploration. Furthermore, we propose a communication approach that, through self-actualization methods, dynamically adapts to the state of the infrastructure. Finally, we offer all of this in an integrated and flexible framework.
\section{Methodology}
In the following, we present our framework for decentralized task offloading in IoT swarms. After presenting the structure of the system we address in our research, we delineate how to let the nodes connect and adjust over time through a semantic and self-actualization-based communication pattern. Furthermore, we present the metaheuristic for the optimization strategy, based on Ant Colony Systems. 

\subsection{Use Case Introduction}
\label{subsec-method:usecase}

\begin{table}[h]
	\centering
	\resizebox{\linewidth}{!}{%
		\begin{tabular}{@{}llllllll@{}}
			\toprule
			\textit{Device type} &
			\textit{Task types} &
			$|\textit{Queue}|$ &
			$C_{comm}$ &
			$C_{comp}$ &
			$C_{storage}$ &
			\textit{P(Fail)} &
			\textit{P(Recover)} \\ \midrule
			$s_1$ & $T_1$, $T_2$        & 10 & 1 & 1 & 1 & 0.1 & 0.4 \\
			$s_2$ & $T_3$, $T_4$, $T_5$ & 15 & 1 & 2 & 1 & 0.1 & 0.4 \\
			$m_1$ & $T_1$, $T_4$        & 10 & 1 & 2 & 2 & 0.2 & 0.3 \\
			$m_2$ & $T_5$               & 5  & 2 & 2 & 2 & 0.2 & 0.3 \\
			$m_3$ & $T_2$, $T_3$        & 10 & 2 & 2 & 3 & 0.2 & 0.3 \\
			$w_1$ & $T_1$               & 5  & 3 & 3 & 3 & 0.3 & 0.2 \\
			$w_2$ & $T_2$               & 5  & 3 & 2 & 3 & 0.3 & 0.2 \\
			$w_3$ & $T_3$               & 5  & 3 & 3 & 3 & 0.3 & 0.2 \\
			$w_4$ & $T_4$               & 5  & 3 & 3 & 3 & 0.3 & 0.2 \\
			$w_5$ & $T_5$               & 5  & 2 & 3 & 3 & 0.3 & 0.2 \\ \bottomrule
		\end{tabular}%
	}
	\caption{Device types and their characteristics.}
	\label{tab:devicetypes}
\end{table}

We expect our framework to work in the case of heterogeneous IoT devices equipped with different processing, storage, and communication capabilities, which serve different kinds of tasks. Each IoT node periodically produces tasks and, in a stable scenario, it takes care of executing them. However, a node may become overloaded and require assistance, rendering task delegation to other swarm nodes crucial. In our use case, the system has three resource groups (computation ($C_{comp}$), communication ($C_{comm}$), and storage ($C_{storage}$)), which can have values between 1 and 3. A \textit{lower value} means that the \textit{device is better} equipped. I.e., the strongest devices in terms of hardware are those where all three values are set to 1. The resources directly influence the processing time of the tasks in the system; Table~\ref{tab:devicetypes} summarizes these properties.  

All the devices in the swarm are connected and can communicate. Therefore, we leverage this scenario to implement autonomous coordination between devices to offload tasks. 
In the development of the framework and of the simulation, we consider the following assumptions: the first is \textbf{heterogeneity}, i.e., the members of an IoT swarm have different hardware specifications (\textit{Device type}), queue sizes ($|\textit{Queue}|$), and emitter types. 
The \textit{Device types} include  strong devices (s), mid-strong devices (m), and resource-constrained
devices, which are the weakest with respect to resources in the swarm (w).
Different IoT devices support different tasks (\textit{Task types}), and the processing time of a task varies depending on the device's capabilities. Furthermore, some devices are more error-prone than others, which affects the reliability of successful task processing.
Secondly, we include \textbf{processing and communication overhead}. Most IoT devices are resource-constrained; therefore, task coordination must avoid negatively impacting the performance of the devices or overflowing the network with messages. Third, \textbf{scalability} is essential. As there is virtually no upper limit to the size of distributed IoT devices in a swarm, successful task coordination algorithms must perform at increasing scales. Finally, our approach is aware of dealing with a \textbf{dynamic environment}. Task coordination must assume that IoT devices might fail or join the system at any time. Hence, coordination must learn the topology and adapt it at runtime.

\subsection{Rain Cloud System (RCS) Architecture}



\begin{figure}
    \centering
    \includegraphics[width=\linewidth]{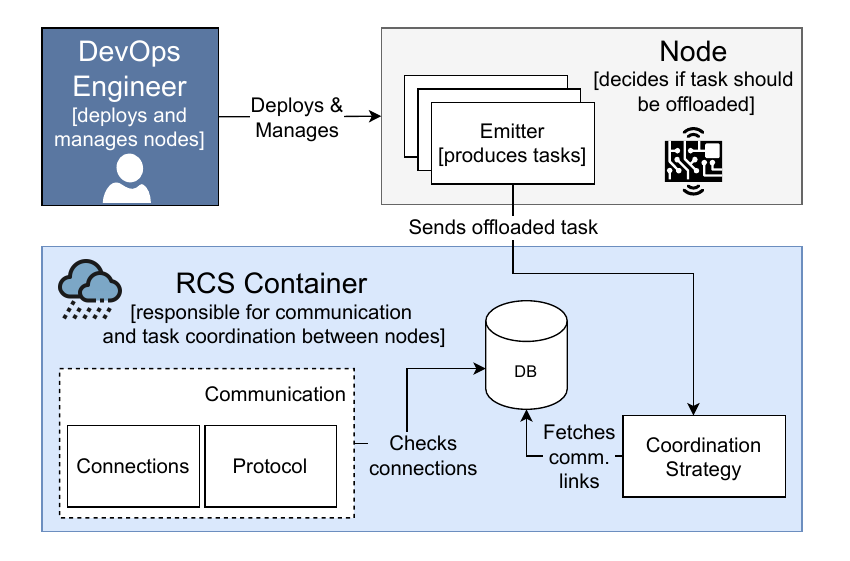}
    \caption{Main components and interaction for the RCS container responsible for a node.}
    \label{fig:RCS-overview}
\end{figure}

Figure~\ref{fig:RCS-overview} depicts the main components and flow for our RCS system, in particular for one node.
In detail, we envision the stakeholder of our framework to be a DevOps engineer, responsible for configuring and managing the IoT devices and the running software.
Therefore, the framework's responsibility is to ensure the machine's connectivity. 
The DevOps engineer manages the IoT swarms by adding new nodes, deleting old ones, or generally specifying connectivity rules, allowing for infrastructure flexibility. 
In this setting, our framework provides task coordination and communication tools.
Tasks are generated by emitters, and each node decides, based on its task queue capacity, whether to process the task locally or forward it to other IoT devices selected according to a coordination strategy.
An example of an IoT device in a smart city is a humidity sensor that provides data to an irrigation application. 
This application example motivated the name of our Rain Cloud System.
After processing this data, the application can use it to determine whether the park's sprinkler system should be activated. 
At this point, the application leverages the framework to communicate with other IoT devices and offload tasks when needed. The application, in turn, transmits and exchanges monitoring data, to evaluate its own state.
To offload a task, the nodes must first connect to other devices in a swarm to exchange their node profiles, which summarize device and task properties.
These node profiles are essential for obtaining information about neighboring devices and new members, which is crucial for task coordination as they reveal the available resources within the swarm.
When a task is offloaded, the application appends a Time-To-Live (TTL) (expressed in seconds) to the request to indicate the maximum waiting time. Once the TTL has expired and no device has been found, the original node executes the task.
The goal is to maintain a balanced task queue, avoiding both underutilization and overutilization and finding a node that can execute an offloaded task with a minimum number of hops.


\subsection{Communication}

The communication module facilitates interaction among devices within the swarm. We establish various protocols to enable efficient intra-swarm communication, employing TCP sockets for message exchange, and JSON as the message format. This interaction is not only essential for exchanging offloading information at runtime, but also for creating and updating the swarm. We handle these two aspects through the Node Discovery and Self-actualization phases.

\subsubsection{Node Discovery}
The Node Discovery (ND) Protocol plays a vital role in the adaptability of the RainCloud System Framework. 
Through the identification and connection of devices, the ND Protocol allows the implementation of a decentralized and interconnected swarm ecosystem.
During the initial handshake phase, nodes exchange profiles. A profile carries a device's crucial information, such as its processing capabilities, memory capacity, available sensors, communication protocols, and supported services. Exchanging profiles during the handshake phase enables information sharing, essential for collaboration and task allocation.
Devices within the swarm can manage the number of open connections they maintain with other members and specify the particular ports they use for establishing these connections. This level of customization ensures the optimization of nodes' performance and resource usage while balancing redundancy and scalability. This flexibility allows the swarm to adapt at runtime according to the environment dynamics.
We design ND so that it relies on semantic information, whereby the protocol draws only the necessary metadata from the devices.\footnote{Semantic information contains metadata as processing capabilities, memory capacity, available sensors, communication protocols, and supported services} This approach enables other protocols to adapt based on the information's meaning, allowing for a more efficient and dynamic data exchange between devices. By focusing on the semantic aspects of device information, the ND Protocol fosters a more intelligent and adaptable IoT ecosystem capable of responding to varying conditions and requirements.
All protocols are designed to work asynchronously.


\subsubsection{Self-actualization process}
Self-actualization (SA) is a concept first introduced by the psychologist Abraham Maslow~\cite{maslow1965self} in the 1950s. It refers to realizing and fulfilling one's potential, representing the highest level of psychological development and the ultimate goal of personal growth.
We use this concept to design a mechanism for dynamically adjusting the swarm topology and enhancing task coordination performance.
The task coordination strategies store a specific value for each outgoing connection to other nodes, which indicates how attractive this connection is. 
For the ACO algorithm, this specific value is the pheromones that are collected over time on a connection variable related to the task type, and for the Gossips algorithm, this value is the number of messages based on the task type for the connection.
SA takes this local knowledge about the swarm and calculates the weakest links. This computation occurs at each node and indicates which links should be disconnected. However, a task might be still offloaded to one of these nodes and, if the connection is terminated, the task might not return. For this reason, the SA process takes place in two phases. In the first stage, the nodes are informed that they should no longer contact specific nodes for offloading tasks because these nodes will soon disconnect. Then, each neighbor stores the detached devices in a block list. Here, connection requests from blocked devices are denied to prevent the same swarm topology from being constructed, forcing disconnected nodes to search for new neighbors. After some time, the marked devices are removed from the list.

\subsection{ACO-inspired offloading mechanism}

\begin{figure}
    \centering
    \includegraphics[width=0.57\linewidth]{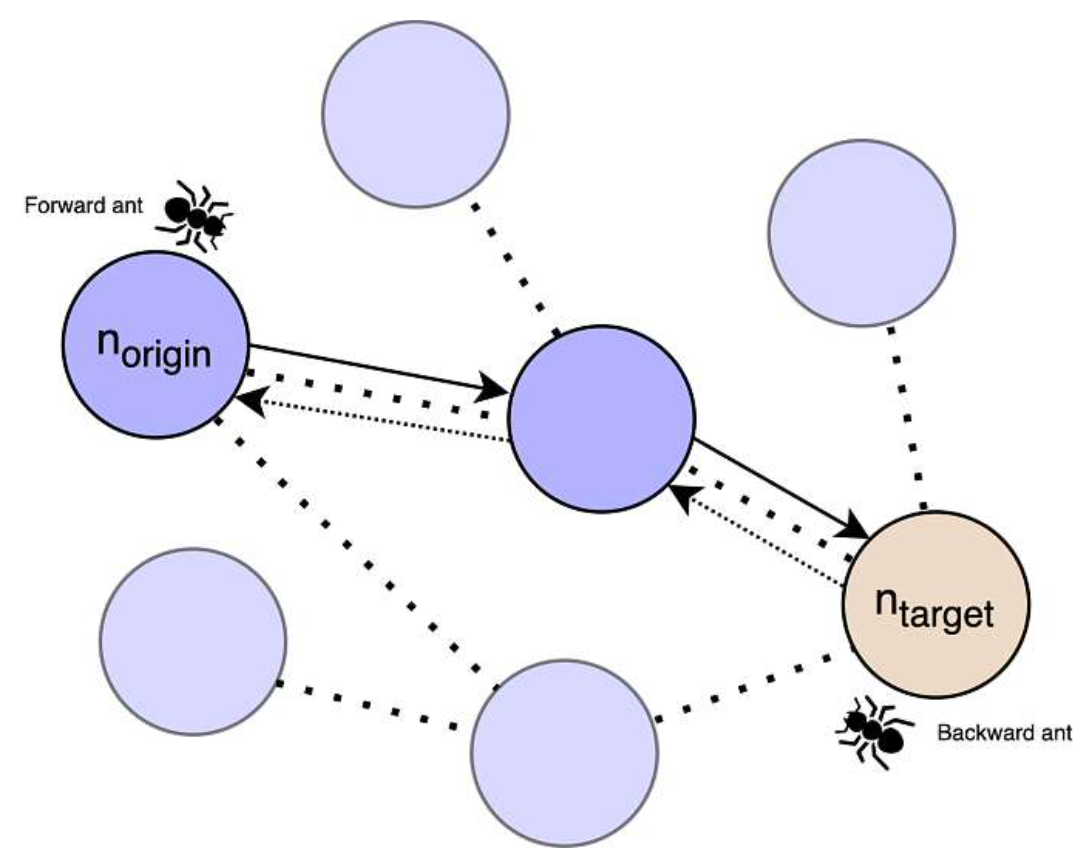}
    \caption{Concept of forward and backward ants in ACO. The forward ant is depicted on the left-hand side, going from the origin to the target, and the backward ant is on the right.}
    \label{fig:aco-ants}
\end{figure}

Here, we explore the proposed metaheuristic for decentralized task offloading. We base our approach on the Ant System (AS), a member of the ACO family of algorithms, bringing in some of its main characteristics for task offloading coordination. Specifically, we aim to minimize the interval between an offloading request and its completion.

\subsubsection{Novelty}%
In our implementation, we first reduce the number of ants to only \textit{a pair} per offload request, exploiting the concept of two different ant agents (see Figure~\ref{fig:aco-ants}), based on~\cite{di1998antnet, michlmayr2006ant}, i.e., the \textit{forward} and \textit{backward} ants. 
%
More in detail, \textit{forward ant} represents the offload request created by the origin device and it is responsible for finding the best path that leads to target nodes. 
Once it finds a suitable node for the task execution along its path, it offloads it there and terminates its role. The \textit{backward ant} retraces the forward ant steps, updating pheromones based on a posteriori knowledge about the node's quality; in our ACO implementation, the \textit{quality} is a function of resource type and availability, plus queue length. We adopt this approach as it avoids the formation of loops~\cite{mohan2012survey} and keeps the process fast and lightweight.
Furthermore, unlike traditional ACO and, to the best of our knowledge, any other approach, our forward ant \textit{carries} the task and TTL. The goal is to minimize message exchange.
Finally, we design the optimization process to last only one iteration, with the goal of improving scalability and reducing computation and communication overhead, accepting some performance loss in task distribution as a trade-off. 

\subsubsection{Detailed Implementation}

%

Formally, in our implementation the transition rule 
is identical to the original AS's approach, i.e., based on a probabilistic decision-making procedure that considers both the pheromone levels and heuristic information associated with the possible transitions through local, a priori knowledge and the global, a posteriori knowledge, but adapted to our scenario. 
Like the traditional approach, we keep the balance between exploration and exploitation by adjusting the influence of these factors through parameters $\alpha$ and $\beta$, where $\alpha$ controls the influence of the a posteriori knowledge and $\beta$ controls the influence of the a priori knowledge.
%
In our case, we define the local knowledge $\eta_{ij}$ as in Equation~\ref{eq:aco-transition-desirability}. It is exchanged during the Node Discovery (ND) protocol and indicates whether two neighboring nodes support the task type in the offload request. If so, a higher pheromone value defined by $\tau_{0+}$ is set. If not, the lower value $\tau_{0}$ is used. The intention is to favor nodes that support the task type to reduce the offloading time and, thus, the latency and number of generated messages.


\begin{equation}
     \eta_{ij}=
    \begin{cases}
      \tau_{0+}, & \text{if next node supports}\ T_{type} \\
      \tau_{0}, & \text{otherwise}
    \end{cases}
    \label{eq:aco-transition-desirability}
\end{equation}

In our approach we express the quality pheromones $Q$ at the target node as the combination of the device's queue quality (Equation~\ref{eq:device-capacities}) and its available capacities (Equation~\ref{eq:quality-pheromones-target}). The device's queue quality $Q_L$ is a function of the average queue occupation, $L_{avg}$, and the queue limit, $L_{limit}$, i.e., when the queue is overloaded. The queue quality is inversely proportional to the queue occupation. The second factor in the quality pheromones calculation is the device's capacities, i.e., communication ($C_{comm}$), computation ($C_{comp}$), and storage ($C_{storage}$).

\begin{equation}
    \text{Q} = Q_L \cdot \frac{1}{\frac{1}{3}(C_{comm} + C_{comp} + C_{storage})}
    \label{eq:quality-pheromones-target}
\end{equation}

\begin{equation}
    Q_L = 
    \begin{cases}
        0, & \text{if}\  1 - \frac{L_{avg}}{L_{limit}} \leq 0 \\
         1 - \frac{L_{avg}}{L_{limit}}, & \text{otherwise}
    \end{cases}
    \label{eq:device-capacities}
\end{equation}

The node capable of processing the task then computes the quality pheromones (see Equation~\ref{eq:quality-pheromones-target}), which the backward ant then uses to update the pheromones in Equation~\ref{eq:aco_updating_pheromones}. The evaporation of the pheromones does not happen as in AS when updating the pheromones but independently every $n$ minutes. 
The backtrack ant updates the pheromones at each node using Equation~\ref{eq:delta_pheromones} that combines the carried quality
pheromones and the distance from the current node to the source node, i.e., $ Q_P = \frac{1}{hopsToOrigin}$.

\begin{equation}
    \tau_{ij} \leftarrow \tau_{ij} + \Delta \tau_{ij}\label{eq:aco_updating_pheromones}
\end{equation}

\begin{equation}
    \Delta \tau_{ij} = 
    (\text{Q} \cdot {Q_P}) \cdot 100
    \label{eq:delta_pheromones}
\end{equation}


With this setting for the quality pheromones, we can capture the device's current availability and inherent capacities, leading to a more informed decision-making process in the ACO algorithm. The strategy aims to select target nodes with higher availability and better resources. Updating the pheromone tables at each node along the path is crucial for learning and guiding the search process. 
Evaporation prevents the algorithm from becoming overly biased towards previous solutions, enabling it to constantly explore new paths and encouraging a balance between exploration and exploitation. Each node in our system performs evaporation independently, removing a relative part of the pheromone values every $n$ minutes. The amount of pheromone reduced is controlled by a parameter $\rho$.

We inspect the time complexity of the ACO-based task coordination strategy, considering different aspects of the algorithm, such as initialization of pheromones, quality pheromones calculation, applying state transition rules, updating of pheromones, and the evaporation process.
\begin{itemize}
    \item \textbf{Pheromones Initialization}. When two devices are connected, the initialization involves updating the pheromone tables for the connected edges. This process has a time complexity of $O(n)$, where $n$ is the number of connected edges.
    \item \textbf{Quality pheromones calculation}. The calculation of quality pheromones at the target node is a constant-time operation, resulting in a time complexity of $O(1)$.
    \item \textbf{State transition rule}. The selection of the next node depends on the state transition rule, which considers the connected edges and the supported task types of the selected node. The time complexity of this process is $O(n \cdot m)$, where $n$ is the number of connected edges and $m$ is the number of supported task types of the selected node.
    \item \textbf{Pheromones update}. Updating pheromones with the backtrack ant involves traversing the connected edges and updating the pheromone values accordingly. This results in a time complexity of $O(n)$, where $n$ is the number of connected edges.
    \item \textbf{Evaporation}. The evaporation process involves updating the pheromone values for each connected edge, and each task type the node has heard of. This results in a time complexity of $O(n \cdot m)$, where $n$ is the number of connected edges and $m$ is the number of task types.
\end{itemize}
Given these adaptations of the AS for our system, we can present the pseudo-code of our ACO strategy. Algorithm~\ref{alg:aco} shows how the ants decide which node to visit next based on a \textit{state transition rule}, which gives a score on the best node. Once the node is selected, the task is computed, and the backward ant updates the pheromones. Finally, periodically, the pheromones evaporate to allow a strategy update.

\begin{algorithm}[h]
    \caption{Optimized Decentralized ACO Strategy}
    \label{alg:aco}
    \DontPrintSemicolon
    \SetAlgoLined
    \KwData{Nodes $nodes$, Task $t$}
    \BlankLine
    \tcp{Concurrent activity at each node}
    \ForEach{node $n$}{
        initializePheromoneTables()\;
        \While{n is running}{
            \If{n is overloaded}{
                createForwardAnt($t$)\;
                \While{$t_{TTL}$ $>$ 0}{
                    $n_j$ = applyStateTransitionRule()\;
                    goToNode($n_j$)\;
                    addNodeToPath($n_j$)\;
                    \If{$n_j$ supports $t_{type}$ and not overloaded}{
                        computeTask($t$)\;
                        createBackwardAnt(calculateQuality(), $path$)\;
                        break\; 
                    }
                }
            }
            \ForEach{backwardAnt}{
                \While{$node \neq node_{origin}$}{
                    $node$ = popNodeFromPath($path$)\;
                    goToNode($node$)\;
                    applyPheromoneTrailUpdateRule()\;
                }
            }
            \If{time for evaporation}{
                applyEvaporationRule()\;
            }
        }
    }
\end{algorithm}

\vspace{-3pt}

\section{Evaluation}
This Section aims to evaluate the performance of the three task coordination algorithms Random, ACO, and Gossips, in the context of decentralized IoT systems. The focus of the evaluation is on considering how the ACO algorithm impacts task coordination’s performance compared to the remaining approaches. To this end, we run tests on servers on which VMs represented as IoT devices run the framework respectively. By running tests with different numbers of devices in each static and dynamic environment, we get a solid overview of how the strategies perform. 
In the static environment, the network topology remains unchanged throughout the experiment.
On the other hand, the dynamic environment simulates decentralized IoT systems’ mobility and ever-changing nature. In this environment, devices can fail and rejoin the network. 
We compare the task coordination implementations through various performance metrics with a focus on the load distribution of the queue, the service time, and the hops per hit.

\vspace{-2pt}

\subsection{Benchmarks}

\paragraph{Random search} The Random strategy is a straightforward, and common~\cite{liu2021reliability, dong2022quantum, chen2024two} technique for decentralized task coordination in distributed IoT systems. Tasks are offloaded to randomly selected nodes without considering their specific characteristics or currently available resources. While its simplicity and ease of implementation guarantee a fast configuration for task offloading, this approach does not possess self-learning or self-adapting capabilities, limiting its ability to optimize task distribution and adapt to dynamic changes in the IoT environment. 
The \textit{computational complexity} of selecting a random neighbor is $O(n)$, where $n$ is the number of neighbors. Thus, the runtime of offloading or forwarding is also $O(n)$.

\paragraph{Gossips}
Furthermore, we design and implement the Gossips approach, adapted from the Gossips protocol. It has two phases: the \textit{Gossips Discovery Phase} and the \textit{Gossips Offload Phase}. The goal of the first phase is to contact as many nodes as possible to find available targets. To this end, the origin node starts spreading messages, so-called Gossips discovery messages, to all its neighbors, which in turn propagate these messages to their neighbors. When a node receives a gossip discovery message and can process the task, it sends a discovery response along the path to the origin node. The first phase continues until the same gossip message is received by a node or the discovery TTL passed, determined by the origin node at the beginning. 
This deterministic approach aims to build extensive knowledge about the system and its available resources, at the same time quickly adapting to changes in the network. In the Gossips Offload Phase the origin node collects all received discovery response messages, which contain information about the current queue capacity and the on-board hardware resources of the target node. Then, this information plus the path length provide a metric for selecting the best target node.
The complexity of the Gossips strategy can be broken down into two main components: spreading and forwarding gossip discovery messages and selecting the target node to offload the task.

\begin{enumerate}
    \item \textbf{Spreading Gossips Discovery Messages}. The complexity of spreading and forwarding Gossips discovery messages can be represented as $O(n)$, where $n$ is the number of connected edges at a node.
    \item \textbf{Selecting the Target Node}. The selection process involves sorting the received response messages and choosing the highest-ranked node as the target for offloading the task. This sorting process has a complexity of $O(n \cdot log(n))$, where $n$ is the number of received gossip discovery response messages.
\end{enumerate}

    \label{fig:arrival_rates}

\subsection{Scenarios}
To thoroughly evaluate the performance of the task coordination strategies, we conducted experiments in two types of environments: \textit{static} and \textit{dynamic}. 
In the \textit{static environment}, the network topology remains unchanged throughout the experiment. This type of environment enables us to assess the performance of the strategies in a stable and controlled setting. On the other hand, in the \textit{dynamic environment}, devices can fail and rejoin the network. Every 15 minutes, we calculate a probability to determine if a node will fail based on each node's characteristics. If the probability is above a certain threshold, the device fails. For each ``failed'' device, every minute, we extract its probability to recover; if it is higher than a threshold, the device rejoins the network. 
These two environments ensure a thorough evaluation of how well the selected strategies can adapt to network changes and maintain performance.
For each environment, we consider settings with different numbers of nodes. The experiments included 10, 25, 50, or 100 devices, sampling both robust nodes and resource-constrained devices (check Section~\ref{subsec-method:usecase}). 
This heterogeneity of devices types and cardinality helps us assess how well the strategies can handle diverse resource capacities in the network. 
For the processing time, we use previously collected application data. 
For the arrival rate of the tasks, we leverage the request generator from the tool Edge Runner\footnote{\url{https://github.com/edgerun/request-generator}}. Each task type is emitted at different times to increase the heterogeneity. $T_1$ and $T_3$ appear most often in the system and are short-lived, while $T_4$ and $T_5$ are the task types emitted least often but with a longer processing time. At any time, an emitter can send out precisely one task. Combined, however, the emitters can produce several tasks at once. Therefore, IoT devices may produce many tasks at once in a short period, resulting in a potential swarm overload that the offloading strategy should handle.
We deploy a total of 100 VMs on the server machines, each of which is equipped with 3072 MB of RAM, 2 vCPU, and 20 GB of disk space.~\footnote{The virtual machines host a Rocky Linux 9.1 OS (minimal installation, \url{https://rockylinux.org/}) and the OpenJDK 11 Java Runtime Environment: OpenJDK 11 (\url{https://openjdk.org/})}

\subsection{Performance metrics}
\label{sec:Performancemetrics}
To effectively compare the performance of the three task coordination strategies, we rely on a set of metrics emanated from related work~\cite{dorigo1997ant, fan2017deadline}. These metrics allow us to analyze various aspects of the strategies and are defined as follows:
\begin{enumerate}
    \item  \textbf{Load Distribution (LD)}. The average queue occupation measures load distribution. The metric reflects the system’s efficiency and aims balanced load, where the queue is neither empty nor too full.
    \item \textbf{Service Time (ST)}. It measures when the origin node sends an offload request for a task until it receives a response back. I.e., how long does the origin have to wait until it can regard the offloaded task as finished? It is expressed in seconds.
    \item \textbf{Hops per Hit (HPH)}. The metric measures the hops required to find the target node. It indicates how often another node must be contacted until the target node is found.
    \item  \textbf{Hit Miss Ratio (HMR)}. The ratio shows how often a target node was found divided by the number of times a target node was not found, but the offload request was lost (e.g., if TTL had already passed).
    \item \textbf{Guarantee Ratio (GR)}. The ratio is calculated by dividing the number of offloaded tasks by the number of returned tasks (from the origin node to the target node and back to the origin node).
    \item \textbf{Amount of Messages (AM)}. The metric counts the number of messages the active strategy produces while coordinating tasks during runtime.
    \item \textbf{Messages per Request (MPR)}. The metric calculates the number of messages produced by the active strategy for a single offload request for a task.
\end{enumerate}

\subsection{Results}
First, we evaluate the three strategies~\footnote{For ACO we have to specify some hyperparameter values. We use the following configuration: $\tau_{0} = 50,\ \tau_{0+}=100,\ \alpha=2,\ \beta=1,\ \rho=0.3$.} by analyzing Load Distribution (LD), Hops-per-Hit (HPH), and Messages per Request (MPR) in a 100-node environment. These selected metrics provide essential insights into the efficiency of the task relocation methods, the frequency of node contacts required to find the target node, and the number of messages generated by a strategy for a single task offload request. Subsequently, we present a comprehensive overview of the three strategies across all metrics described in the previous Subsection \ref{sec:Performancemetrics} to offer a broader perspective on performance.

\subsubsection{Load Distribution (LD)}
The primary aim of LD is to assess the balancing of load across the system. A balanced LD ensures that tasks are offloaded to appropriate target nodes, preventing overload on specific nodes while others remain underutilized.
We use the term queue capacity to describe how many tasks must be in the queue before subsequent tasks must be offloaded. Queue occupation describes the total number of tasks in the queue, including both local tasks and tasks that will be offloaded. If the queue occupation exceeds the capacity, we call the device overloaded. The average queue occupation in our system is the LD described above. 
We experiment with a varying number of nodes, in both static and dynamic settings. For the sake of brevity, we show in Figure~\ref{fig:ld-100n} only the results for the case of 100 nodes. The rest of the values can be found in the summarizing tables in Section~\ref{sec:summary}.
In Figure~\ref{fig:ld-100n}, we can observe how the static Random approach exhibits stronger deviations from the regression line and the queue occupation grows at a queue capacity of ten. With a queue capacity of 15, this trend is amplified: Random's task coordination loses performance, and the prediction states that the devices are overloaded after one hour. Figure~\ref{fig:ld-100n-boxplot} emphasises this behavior. Especially for small queue sizes, the three strategies tend to have requests that overflow the queue size, calling for more accurate offloading mechanisms. This behavior is mitigated by larger queue sizes for ACO and Gossips, whereas Random Search still shows irregularities. We conclude that Random can only cope with smaller queue capacities for large swarm sizes. Gossips shows the smallest LD values, while ACO has the highest values but never overloads the devices, which can be concluded as a solid task coordination mechanism.

In summary, ACO demonstrates the most effective utilization of queue resources across various scenarios and swarm sizes, likely due to its self-learning nature. Random shows that with large swarm sizes and larger queue capacities, it overloads the devices and thus degrades the overall system performance. The Gossips strategy shows the most stable values without the risk of overloading devices, but the queue utilization is in the lower range. In general, all three approaches have underutilization at the five and ten queue capacities, which can be improved in future works.

\begin{figure}[ht!]
  \centering
  \begin{subfigure}[b]{0.45\linewidth}
    \includegraphics[width=\linewidth]{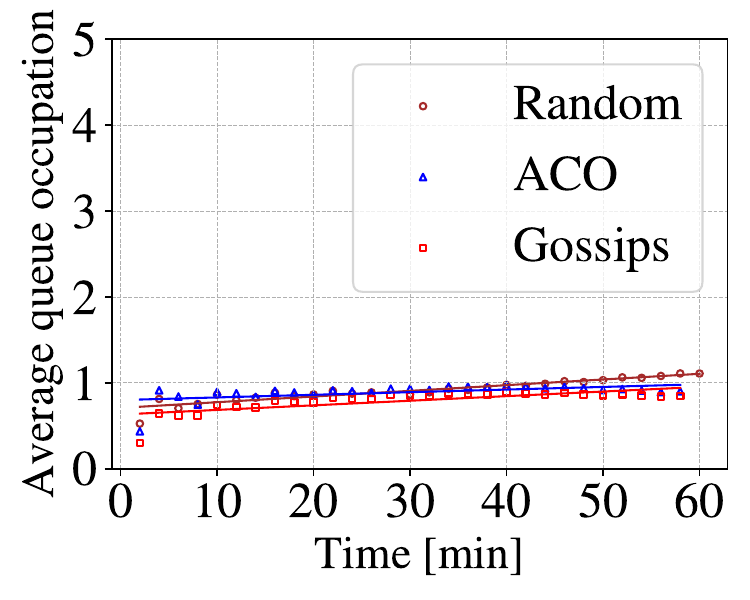}
    \caption{$|\text{Queue}| = 5$. Static.}
  \end{subfigure}
  \hfill
  \begin{subfigure}[b]{0.45\linewidth}
    \includegraphics[width=\linewidth]{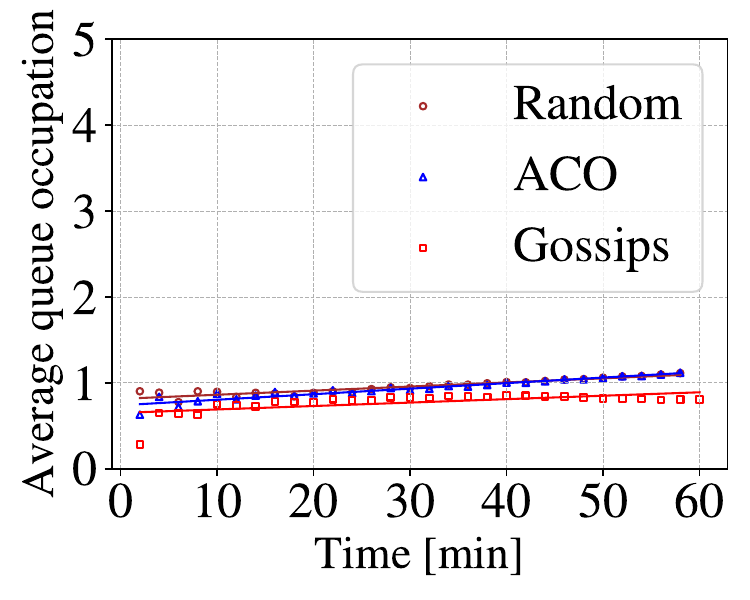}
    \caption{$|\text{Queue}| = 5$. Dynamic.}
  \end{subfigure}
  \\
  \begin{subfigure}[b]{0.45\linewidth}
    \includegraphics[width=\linewidth]{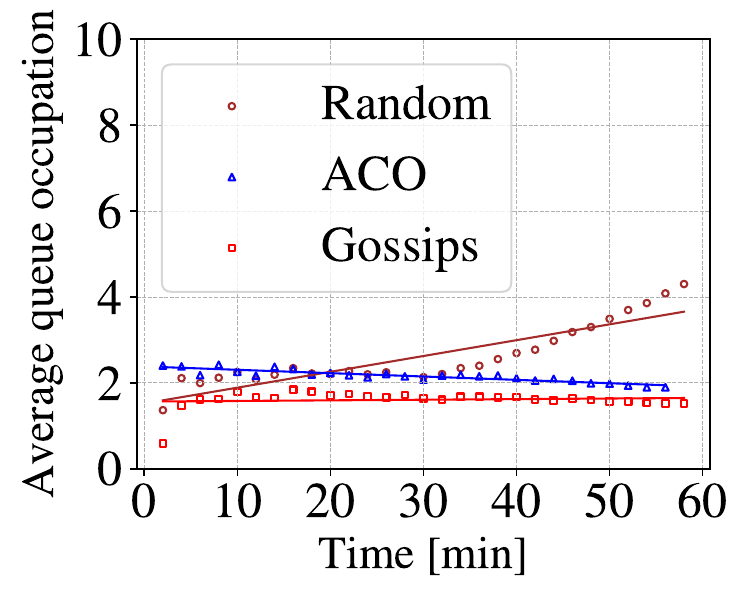}
    \caption{$|\text{Queue}| = 10$. Static.}
  \end{subfigure}
  \hfill
  \begin{subfigure}[b]{0.45\linewidth}
    \includegraphics[width=\linewidth]{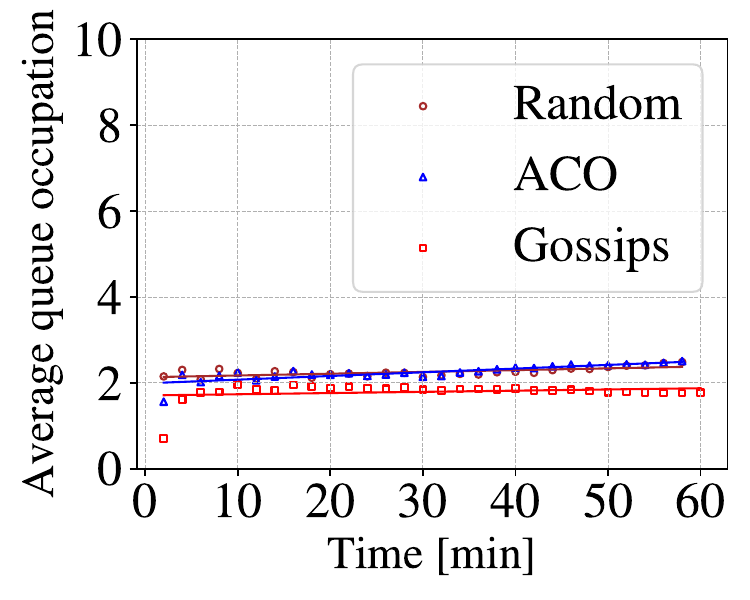}
    \caption{$|\text{Queue}|= 10$. Dynamic.}
  \end{subfigure}
  \\
  \begin{subfigure}[b]{0.45\linewidth}
    \includegraphics[width=\linewidth]{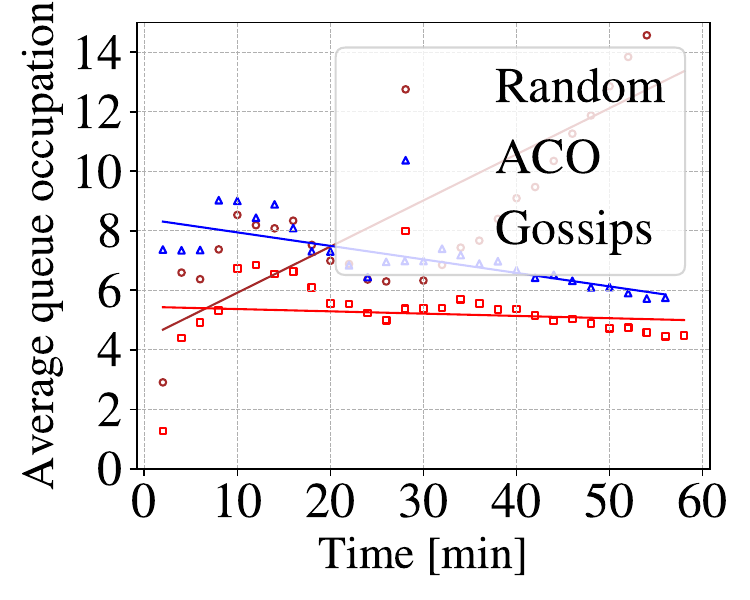}
    \caption{$|\text{Queue}| = 15$. Static.}
  \end{subfigure}
  \hfill
  \begin{subfigure}[b]{0.45\linewidth}
    \includegraphics[width=\linewidth]{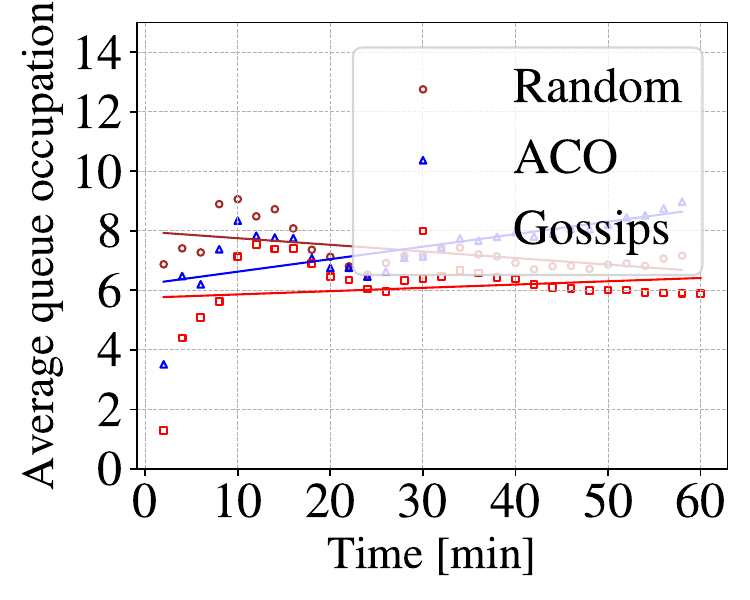}
    \caption{$|\text{Queue}| = 15$. Dynamic.}
  \end{subfigure}
  \caption{LD in static (1\textsuperscript{st} column) and dynamic (2\textsuperscript{nd} column) environments with 100 devices of different queue capacities.}
  \label{fig:ld-100n}
\end{figure}

\begin{figure}[ht!]
  \centering
  \begin{subfigure}[b]{0.45\linewidth}
    \includegraphics[width=\linewidth]{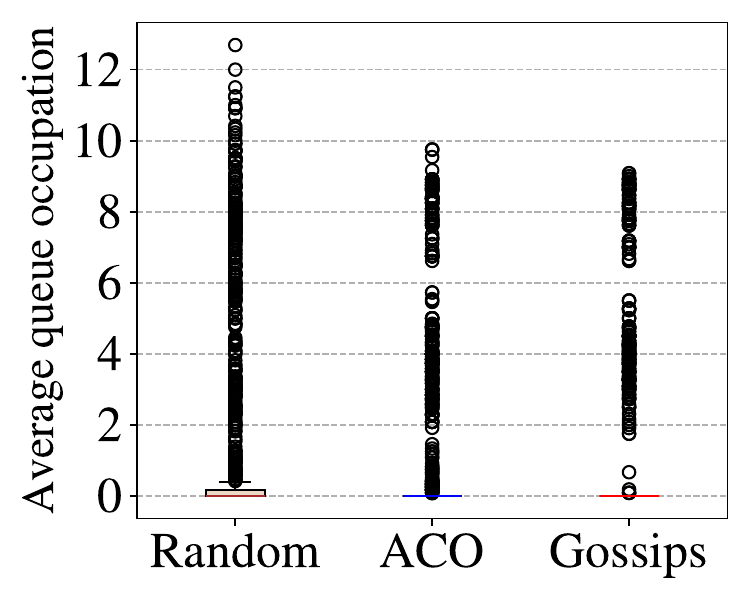}
    \caption{$|\text{Queue}| = 5$. Static.}
  \end{subfigure}
  \hfill
  \begin{subfigure}[b]{0.45\linewidth}
    \includegraphics[width=\linewidth]{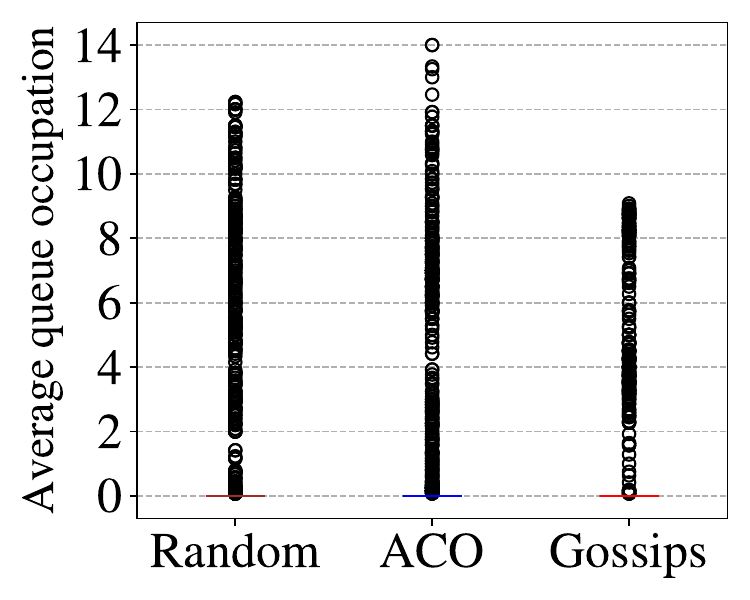}
    \caption{$|\text{Queue}| = 5$. Dynamic.}
  \end{subfigure}
  \\
  \begin{subfigure}[b]{0.45\linewidth}
    \includegraphics[width=\linewidth]{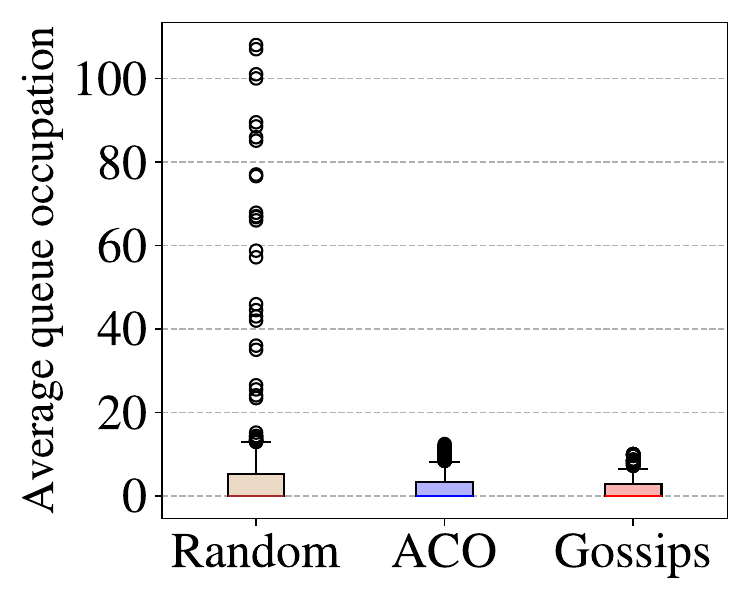}
    \caption{$|\text{Queue}| = 10$. Static.}
  \end{subfigure}
  \hfill
  \begin{subfigure}[b]{0.45\linewidth}
    \includegraphics[width=\linewidth]{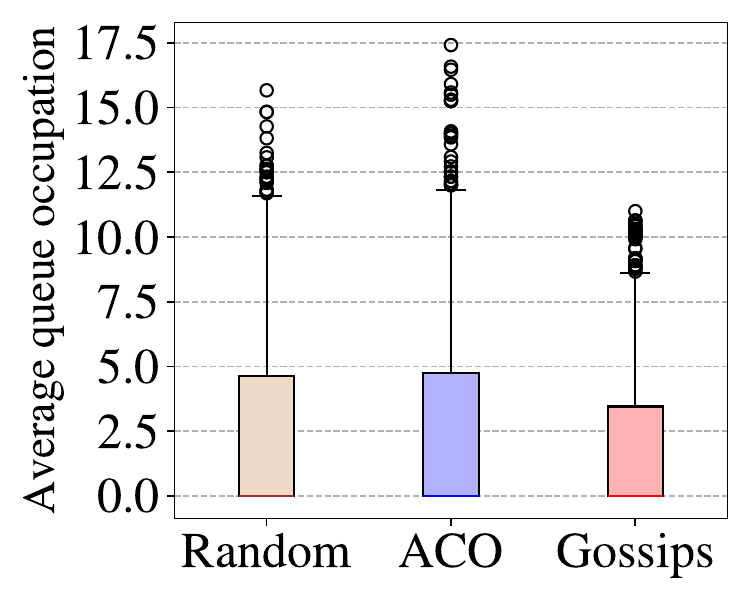}
    \caption{$|\text{Queue}|= 10$. Dynamic.}
  \end{subfigure}
  \\
  \begin{subfigure}[b]{0.45\linewidth}
    \includegraphics[width=\linewidth]{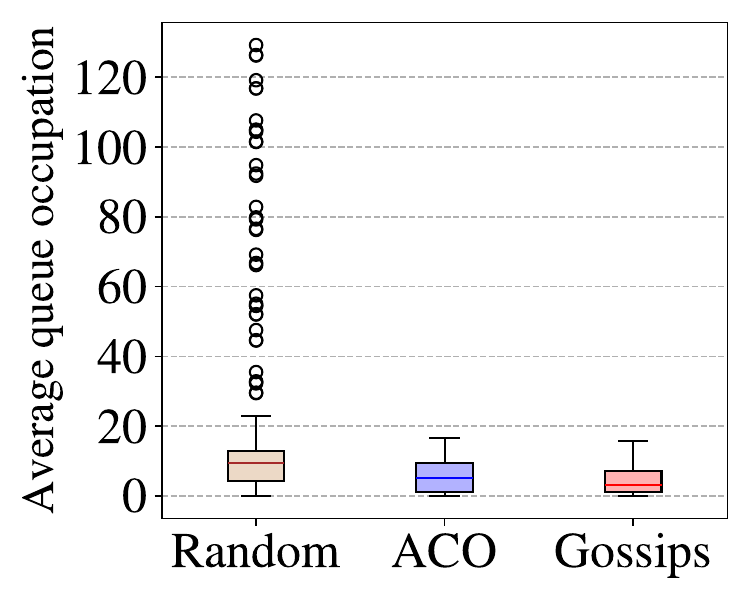}
    \caption{$|\text{Queue}| = 15$. Static.}
  \end{subfigure}
  \hfill
  \begin{subfigure}[b]{0.45\linewidth}
    \includegraphics[width=\linewidth]{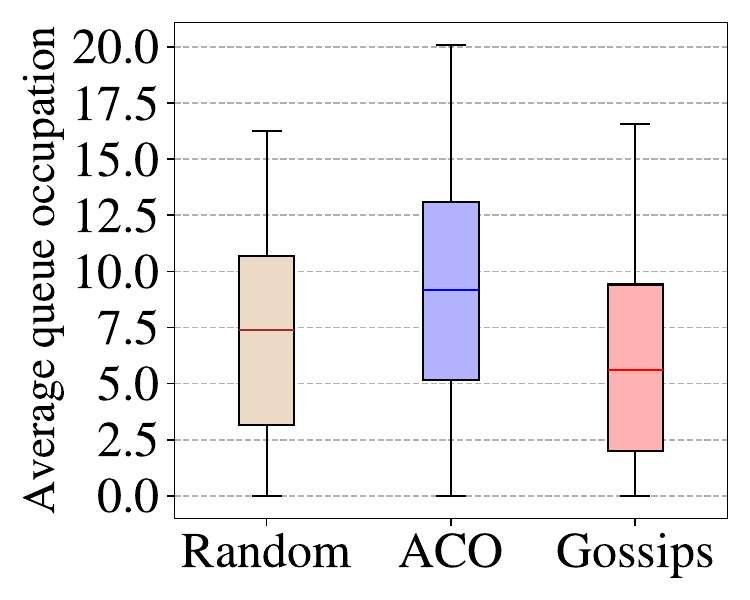}
    \caption{$|\text{Queue}| = 15$. Dynamic.}
  \end{subfigure}
  \caption{LD in static (1\textsuperscript{st} column) and dynamic (2\textsuperscript{nd} column) environments with 100 devices of different queue capacities.}
  \label{fig:ld-100n-boxplot}
\end{figure}

\subsubsection{Hops-per-Hit (HPH)}
\begin{figure}[h]
	\centering
	\begin{subfigure}[b]{0.45\linewidth}
		\includegraphics[width=\textwidth]{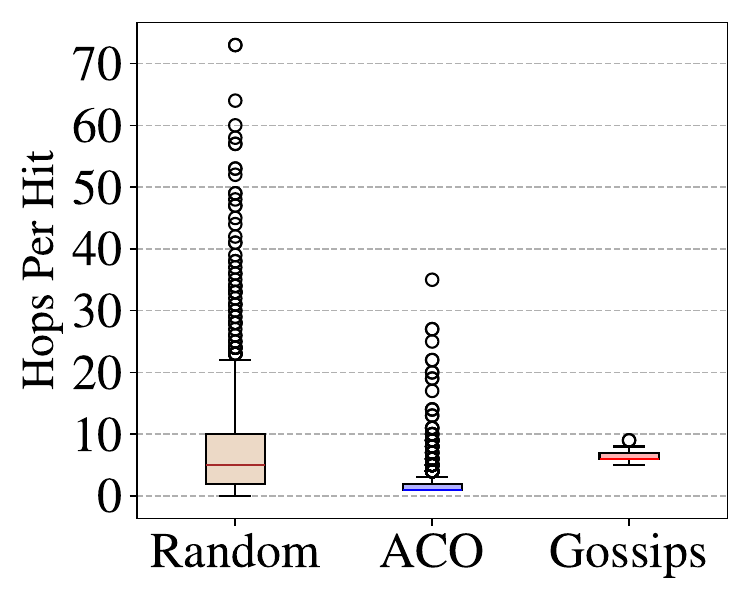}
		\caption{Static HPH.}
	\end{subfigure}
	\hfill
	\begin{subfigure}[b]{0.45\linewidth}
		\includegraphics[width=\textwidth]{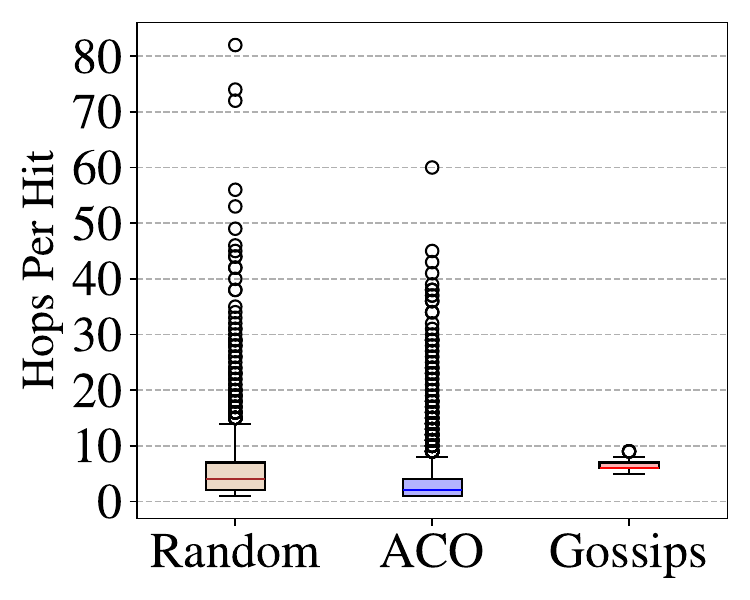}
		\caption{Dynamic HPH.}
	\end{subfigure}
	\caption{HPH in static and dynamic environments with 100 devices.}
	\label{fig:hph-100}
\end{figure}

\vspace{-2pt}

Assessing the average HPH metric allows us to understand the efficiency and scalability of the Random, ACO, and Gossips strategies in the context of message propagation and task coordination.
HPH measures the number of hops or nodes between the origin and target nodes during task offloading. 
A lower HPH value indicates fewer messages sent and a closer target node, leading to reduced latency, faster communication, and shorter service times.
Figure~\ref{fig:hph-100} summarizes the test run with 100 nodes. Given a swarm size of 100, Random is the worst algorithm in terms of HPH, while ACO has the best mean value. Compared to the Gossips, Random and ACO have more outliers. In a static environment, Random's highest value is even 75, meaning one-third of all nodes were traversed until the task offload request was accepted. Gossips shows stable and consistent values in both environments.
In summary, the ACO strategy has the lowest and, thus, best HPH values in both environment setups, thanks to the deposition of pheromones. 
It would be interesting to know if the outliers are caused by the fact that the system needs more time to learn due to the larger number of nodes. I.e., the long paths (outliers) are formed at the beginning because not enough artificial ants have explored the network. Gossips scores well in this category and shows solid stability and consistency in static and dynamic environments, while Random has the worst performance.

\begin{figure}[h]
	\centering
    \begin{subfigure}[b]{0.49\linewidth}
		\begin{tikzpicture}
			\begin{axis}[
					ymode=log,
                    xlabel={Time [min]},
					ylabel={\#Messages per request},
					ymajorgrids=true,
					grid style=dashed,
					width=1.0\linewidth,
                    height=1.0\linewidth,
                    legend to name=sharedMPRLegend,
                    legend style={at={(0.5,-0.38)}, anchor=north,legend columns=3},
				]
				\addplot[line width=0.2mm, color=brown] table[x=time, y=value,col sep=comma] 
				{Figures/data_for_plots/messages-per-request/rw-n100-time.csv};

				\addplot[line width=0.2mm, color=blue] table[x=time, y=value,col sep=comma]
				{Figures/data_for_plots/messages-per-request/aco-n100-time.csv};

                \addplot[line width=0.2mm, color=red] table[x=time, y=value,col sep=comma]
				{Figures/data_for_plots/messages-per-request/gossips-n100-time.csv};

                \addplot[line width=0.2mm, color=brown, dashed] table[x=time, y=value,col sep=comma]
				{Figures/data_for_plots/messages-per-request/rw-n100-time-average.csv};
    
				\addplot[line width=0.2mm, color=blue, dashed] table[x=time, y=value,col sep=comma]
				{Figures/data_for_plots/messages-per-request/aco-n100-time-average.csv};
								            
				\addplot[line width=0.2mm, color=red, dashed] table[x=time, y=value,col sep=comma]
				{Figures/data_for_plots/messages-per-request/gossips-n100-time-average.csv};
    
				\addlegendentry{Random}
                \addlegendentry{ACO}
                \addlegendentry{Gossip}
				\addlegendentry{Random Avg}
                \addlegendentry{ACO Avg}
                \addlegendentry{Gossip Avg}
			\end{axis}
        \end{tikzpicture}
    \end{subfigure} \hfill
    \begin{subfigure}[b]{0.49\linewidth}
        \begin{tikzpicture}
			\begin{axis}[
					ymode=log,
                    xlabel={Time [min]},
					ylabel={\#Messages per request},
					ymajorgrids=true,
					grid style=dashed,
					width=1.0\linewidth,
                    height=1.0\linewidth,
                    legend to name=sharedMPRLegend,
                    legend style={at={(0.5,-0.38)}, anchor=north,legend columns=3},
				]
				\addplot[line width=0.2mm, color=brown] table[x=time, y=value,col sep=comma] 
				{Figures/data_for_plots/messages-per-request/rw-n100-time-fail.csv};
											
				\addplot[line width=0.2mm, color=blue] table[x=time, y=value,col sep=comma]
				{Figures/data_for_plots/messages-per-request/aco-n100-time-fail.csv};

                \addplot[line width=0.2mm, color=red] table[x=time, y=value,col sep=comma]
				{Figures/data_for_plots/messages-per-request/gossips-n100-time-fail.csv};

                \addplot[line width=0.2mm, color=brown, dashed] table[x=time, y=value,col sep=comma]
				{Figures/data_for_plots/messages-per-request/rw-n100-time-fail-average.csv};
    
				\addplot[line width=0.2mm, color=blue, dashed] table[x=time, y=value,col sep=comma]
				{Figures/data_for_plots/messages-per-request/aco-n100-time-fail-average.csv};
								            
				\addplot[line width=0.2mm, color=red, dashed] table[x=time, y=value,col sep=comma]
				{Figures/data_for_plots/messages-per-request/gossips-n100-time-fail-average.csv};
    
				\addlegendentry{Random}
                \addlegendentry{ACO}
                \addlegendentry{Gossip}
				\addlegendentry{Random Avg}
                \addlegendentry{ACO Avg}
                \addlegendentry{Gossip Avg}
                \end{axis}
            \end{tikzpicture}
        \end{subfigure}
  \ref{sharedMPRLegend}

  \caption{Message per request over time in \textit{static} (\textit{left}) and \textit{dynamic} (\textit{right}) environments with 100 devices.}
  \label{fig:mpr-100-time}
\end{figure}
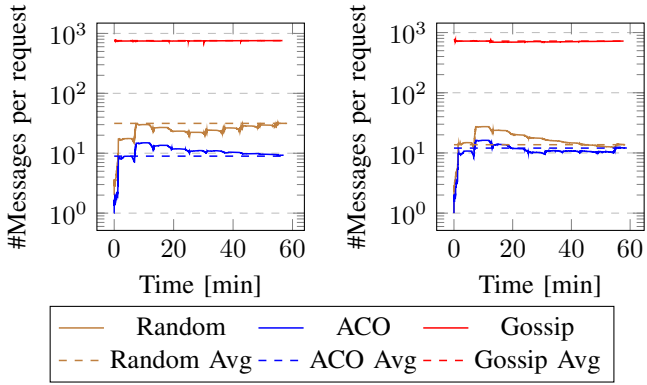
\subsubsection{Messages per requests (MPR)}
The MPR metric allows us to analyze the typical workload in RCS and to process its ability to deliver tasks to its destination.
In Figure~\ref{fig:mpr-100-time} we can analyze the results with a setup of 100 nodes. In this scenario, the ACO algorithm demonstrates its effectiveness in managing communication efficiency within distributed systems. In contrast, the Gossips algorithm exhibits instability in messages per request regard, and the Random plateaus at higher values.

\subsubsection{Summary}
\label{sec:summary}
An overview of the performance of the strategies in a static environment is provided in Table~\ref{tab:static_environment}, while the results for the dynamic environment are presented in Table~\ref{tab:dynamic_environment}.
They both report the average values for each metric, truncated after two comma digits.
The winning strategy in the metric and swarm size category is highlighted in bold and green; the second place is colored orange, and the last place is red. 
For LD, weighting is based on whether the queue is neither too full nor too empty – strategies should target the occupation of 3/4 of the queue, given a capacity of 15.
Lower values indicate better performance for ST, HPH, AM, and MPR. 
For the two metrics, HMR and GR, higher scores indicate better performance by the strategies.

Overall, the proposed ACO algorithm shows great potential for successfully coordinating nodes in a fully decentralized way. The proposed solution constantly shows the best performance in many of the used metrics. Random outperforms only slightly ACO on service time but has poor performance in terms of actual coordination, whereas Gossip shows better results for HMR and GR but requires much more Service Time and sends in the systems a much higher number of messages, with a serious risk of clogging the network. 
Furthermore, the test has only run for one hour; longer training times for ACO will intuitively improve its performance in HMR and GR, due to better exploration and refined pheromone trails, with the potential of making it solidly the preferred solution.

\begin{table}[htbp]
\resizebox{0.5\textwidth}{!}{
\begin{tabular}{lllclcccc}
		\hline
		\textbf{Strategy}          &   
		\textbf{\#Nodes}           &   
		\textbf{LD}                &   
		\textbf{ST}                &   
		\textbf{HPH}               &   
		\textbf{AM}                &   
		\textbf{MPR}               &   
		\textbf{HMR}               &   
		\textbf{GR} \\ \hline
		Random                     &   
		10                         &   
		\cellcolor{red!25}4.46     &   
		\cellcolor{orange!25}64.06 &   
		\cellcolor{red!25}4.28     &   
		\cellcolor{orange!25}713   &   
		\cellcolor{orange!25}5.40  &   
		\cellcolor{red!25}45.45    &   
		\cellcolor{red!25}45.45 \\
		                           &   
		25                         &   
		\cellcolor{orange!25}4.97  &   
		\cellcolor{green!25}\textbf{65.85}  &   
		\cellcolor{orange!25}4.71  &   
		\cellcolor{orange!25}3215  &   
		\cellcolor{orange!25}7.29  &   
		\cellcolor{red!25}48.29    &   
		\cellcolor{red!25}48.29 \\
		                           &   
		50                         &   
		\cellcolor{orange!25}5.14  &   
		\cellcolor{green!25}\textbf{66.80}  &   
		\cellcolor{red!25}6.20     &   
		\cellcolor{orange!25}8865  &   
		\cellcolor{orange!25}11.40 &   
		\cellcolor{red!25}46.07    &   
		\cellcolor{red!25}46.07 \\
		                           &   
		100                        &   
		\cellcolor{red!25}16.73    &   
		\cellcolor{orange!25}72.21 &   
		\cellcolor{red!25}7.82     &   
		\cellcolor{orange!25}93818 &   
		\cellcolor{orange!25}31.50 &   
		\cellcolor{red!25}35.75    &   
		\cellcolor{red!25}35.65 \\ \hline
		ACO                        &   
		10                         &   
		\cellcolor{green!25}\textbf{5.03}   &   
		\cellcolor{green!25}\textbf{62.79}  &   
		\cellcolor{green!25}\textbf{1.34}   &   
		\cellcolor{green!25}\textbf{300}    &   
		\cellcolor{green!25}\textbf{2.37}   &   
		\cellcolor{orange!25}57.26 &   
		\cellcolor{orange!25}57.26 \\
		                           &   
		25                         &   
		\cellcolor{green!25}\textbf{5.53}   &   
		\cellcolor{orange!25}70.26 &   
		\cellcolor{green!25}\textbf{2.01}   &   
		\cellcolor{green!25}\textbf{2309}   &   
		\cellcolor{green!25}\textbf{4.75}   &   
		\cellcolor{orange!25}50.84 &   
		\cellcolor{orange!25}50.84 \\
		                           &   
		50                         &   
		\cellcolor{green!25}\textbf{5.40}   &   
		\cellcolor{orange!25}71.27 &   
		\cellcolor{green!25}\textbf{1.95}   &   
		\cellcolor{green!25}\textbf{4664}   &   
		\cellcolor{green!25}\textbf{5.75}   &   
		\cellcolor{orange!25}47.15 &   
		\cellcolor{orange!25}47.15 \\
		                           &   
		100                        &   
		\cellcolor{green!25}\textbf{5.75}   &   
		\cellcolor{green!25}\textbf{71.19}  &   
		\cellcolor{green!25}\textbf{2.38}   &   
		\cellcolor{green!25}\textbf{14243}  &   
		\cellcolor{green!25}\textbf{8.93}   &   
		\cellcolor{orange!25}48.69 &   
		\cellcolor{orange!25}48.69 \\ \hline
		Gossips                    &   
		10                         &   
		\cellcolor{orange!25}4.60  &   
		\cellcolor{red!25}106.35   &   
		\cellcolor{orange!25}3.44  &   
		\cellcolor{red!25}3900     &   
		\cellcolor{red!25}33.42    &   
		\cellcolor{green!25}\textbf{69.36}  &   
		\cellcolor{green!25}\textbf{69.36} \\
		                           &   
		25                         &   
		\cellcolor{red!25}4.58     &   
		\cellcolor{red!25}107.34   &   
		\cellcolor{red!25}5.09     &   
		\cellcolor{red!25}37400    &   
		\cellcolor{red!25}114.01   &   
		\cellcolor{green!25}\textbf{58.07}  &   
		\cellcolor{green!25}\textbf{58.07} \\
		                           &   
		50                         &   
		\cellcolor{red!25}4.38     &   
		\cellcolor{red!25}108.14   &   
		\cellcolor{orange!25}5.97  &   
		\cellcolor{red!25}163549   &   
		\cellcolor{red!25}296.12   &   
		\cellcolor{green!25}\textbf{53.11}  &   
		\cellcolor{green!25}\textbf{53.11} \\
		                           &   
		100                        &   
		\cellcolor{orange!25}4.47  &   
		\cellcolor{red!25}103.68   &   
		\cellcolor{orange!25}6.42  &   
		\cellcolor{red!25}827354   &   
		\cellcolor{red!25}757.37   &   
		\cellcolor{green!25}\textbf{58.02}  &   
		\cellcolor{green!25}\textbf{58.02} \\ \hline
	\end{tabular}}
	\caption{Comparison of strategies in a static environment.}
	\label{tab:static_environment}
\end{table}
\begin{table}[htbp]
\resizebox{0.5\textwidth}{!}{
    \begin{tabular}{lllclcccc}
    \toprule
		\textbf{Strategy}          &   
		\textbf{\#Nodes}           &   
		\textbf{LD}                &   
		\textbf{ST}                &   
		\textbf{HPH}               &   
		\textbf{AM}                &   
		\textbf{MPR}               &   
		\textbf{HMR}               &   
		\textbf{GR} \\ \hline
		Random                     &   
		10                         &   
		\cellcolor{green!25}\textbf{7.35}   &   
		\cellcolor{green!25}\textbf{60.39}  &   
		\cellcolor{orange!25}3.34  &   
		\cellcolor{orange!25}1082  &   
		\cellcolor{orange!25}5.07  &   
		\cellcolor{red!25}43.19    &   
		\cellcolor{red!25}43.19 \\
		                           &   
		25                         &   
		\cellcolor{orange!25}9.47  &   
		\cellcolor{green!25}\textbf{76.15}  &   
		\cellcolor{orange!25}4.22  &   
		\cellcolor{green!25}\textbf{8220}   &   
		\cellcolor{orange!25}8.40  &   
		\cellcolor{orange!25}42.33 &   
		\cellcolor{orange!25}41.30 \\
		                           &   
		50                         &   
		\cellcolor{green!25}\textbf{8.62}   &   
		\cellcolor{green!25}\textbf{73.09}  &   
		\cellcolor{red!25}6.90     &   
		\cellcolor{orange!25}20100 &   
		\cellcolor{orange!25}13.74 &   
		\cellcolor{red!25}45.55    &   
		\cellcolor{red!25}44.39 \\
		                           &   
		100                        &   
		\cellcolor{red!25}7.16     &   
		\cellcolor{green!25}\textbf{71.37}  &   
		\cellcolor{orange!25}6.15  &   
		\cellcolor{green!25}\textbf{30211}  &   
		\cellcolor{orange!25}13.70 &   
		\cellcolor{orange!25}48.79 &   
		\cellcolor{orange!25}47.02 \\ \hline
		ACO                        &   
		10                         &   
		\cellcolor{orange!25}4.97  &   
		\cellcolor{orange!25}63.78 &   
		\cellcolor{green!25}\textbf{1.11}   &   
		\cellcolor{green!25}\textbf{295}    &   
		\cellcolor{green!25}\textbf{2.36}   &   
		\cellcolor{orange!25}52.94 &   
		\cellcolor{orange!25}52.94 \\
		                           &   
		25                         &   
		\cellcolor{green!25}\textbf{11.48}  &   
		\cellcolor{orange!25}85.80 &   
		\cellcolor{green!25}\textbf{3.36}   &   
		\cellcolor{orange!25}8354  &   
		\cellcolor{green!25}\textbf{7.99}   &   
		\cellcolor{red!25}40.22    &   
		\cellcolor{red!25}38.79 \\
		                           &   
		50                         &   
		\cellcolor{orange!25}6.30  &   
		\cellcolor{orange!25}70.87 &   
		\cellcolor{green!25}\textbf{2.85}   &   
		\cellcolor{green!25}\textbf{6693}   &   
		\cellcolor{green!25}\textbf{7.00}   &   
		\cellcolor{orange!25}47.06 &   
		\cellcolor{orange!25}46.75 \\
		                           &   
		100                        &   
		\cellcolor{green!25}\textbf{8.97}   &   
		\cellcolor{orange!25}78.97 &   
		\cellcolor{green!25}\textbf{4.23}   &   
		\cellcolor{orange!25}32160 &   
		\cellcolor{green!25}\textbf{12.03}  &   
		\cellcolor{red!25}47.64    &   
		\cellcolor{red!25}46.44 \\ \hline
		Gossips                    &   
		10                         &   
		\cellcolor{red!25}3.69     &   
		\cellcolor{red!25}91.27    &   
		\cellcolor{red!25}3.69     &   
		\cellcolor{red!25}3259     &   
		\cellcolor{red!25}32.47    &   
		\cellcolor{green!25}\textbf{78.16}  &   
		\cellcolor{green!25}\textbf{74.71} \\
		                           &   
		25                         &   
		\cellcolor{red!25}5.10     &   
		\cellcolor{red!25}106.92   &   
		\cellcolor{red!25}5.03     &   
		\cellcolor{red!25}43816    &   
		\cellcolor{red!25}121.22   &   
		\cellcolor{green!25}\textbf{58.60}  &   
		\cellcolor{green!25}\textbf{58.60} \\
		                           &   
		50                         &   
		\cellcolor{red!25}5.92     &   
		\cellcolor{red!25}104.56   &   
		\cellcolor{orange!25}5.56  &   
		\cellcolor{red!25}196590   &   
		\cellcolor{red!25}304.28   &   
		\cellcolor{green!25}\textbf{51.20}  &   
		\cellcolor{green!25}\textbf{51.04} \\
		                           &   
		100                        &   
		\cellcolor{orange!25}5.88  &   
		\cellcolor{red!25}110.33   &   
		\cellcolor{red!25}6.50     &   
		\cellcolor{red!25}849374   &   
		\cellcolor{red!25}720.20   &   
		\cellcolor{green!25}\textbf{52.32}  &   
		\cellcolor{green!25}\textbf{52.23} \\ \hline
	\end{tabular}
}
\caption{Comparison of strategies in a dynamic environment.}
	\label{tab:dynamic_environment}
\end{table}

\vspace{-2pt}

\section{Conclusion}
\label{sec:conclusion}
This paper introduced the Rain Cloud System (RCS) framework to facilitate decentralized swarm task offloading.
In particular, we developed and evaluated a semantic and self-actualizing communication strategy, coupled with a metaheuristic based on the Ant Colony System and compared it with Random Search and Gossip, adapted for the use case. 
The Ant Colony Optimization (ACO) algorithm demonstrated promising results in network management and resource distribution, indicating its suitability for complex IoT challenges. ACO and the other algorithms embedded in the RCS framework allow the development of inventive strategies for resource distribution and task management within IoT networks.

In future work, we will incorporate path optimization for ACO, avoiding cycles.
Furthermore, we aim to optimize the pheromones deposit during the initialization process, using more accurate rules. In general, the pheromone calculation could also take into account energy consumption or device geographical proximity. In addition, we want to explore a hybrid solution that combines the strengths of both ACO and Gossips strategies, as the latter performs better in dynamic settings.
Lastly, we aim at performing additional experiments as, for example, an exhaustive parameter testing for ACO.
\section*{Acknowledgment}
This work is funded by the HORIZON Research and Innovation Action 101135576 INTEND ``Intent-based data operation in the computing continuum.''

\bibliographystyle{IEEEtran}
\bibliography{IEEEabrv, references}

\end{document}